\documentclass[%
 reprint,
 amsmath,amssymb,
 aps,
 upgreek,
]{revtex4-2}
\usepackage{chemformula}
\usepackage{upgreek}
\usepackage[T1]{fontenc}
\usepackage{graphicx}
\usepackage{dcolumn}
\usepackage{bm}

\begin{document}

\preprint{APS/123-QED}
\title{Superfluidic nature of self-driven nanofluidics at liquid-gas interfaces}

\author{Vinitha Johny$^{1,2}$}
\author{Sonia Contera$^3$}
\author{Siddharth Ghosh$^{1,2,4}$}%
 \email{Email of correspondence: sg915@cam.ac.uk}
\affiliation{%
$^1$International Center for Nanodevices, Center for Nano Science and Engineering, Indian Institute of Science, Bangalore.\\
$^2$Department of Science, Open Academic Research Council, Kolkata, India \& Department of Science, Open Academic Research UK CIC, Cambridge, UK.\\
$^3$Department of Physics, University of Oxford, UK.\\
$^4$Department of Applied Mathematics and Theoretical Physics, University of Cambridge, UK.
}%


\begin{abstract}
Self-driven nanofluidic flow at the liquid-air interface is a non-intuitive phenomenon. This flow behaviour was not driven by classical pressure difference or evaporation only. Depending on the position of the nanofluidic pore we can observe flow and no-flow with chaotic behaviour. In this paper, we study the nonlinear dynamics of a confined nanopore system at the liquid-air interface. The finite-range interactions between the interacting species are quantified with a corresponding critical velocity of the system. This is visualised using the finite element method and analysed mathematically with the Landau criterion. We found the formation of Bose-Einstein-like condensates due to the  transport through nanofluidic pores. We show that systems with more than one nanofluidic pore with a sub-100 nm diameter create a highly nonlinear and complex. The approximation of relevant classical systems to existing quantum mechanical systems divulges new results and corrections at a fundamental level. We explain the formation of oscillating condensate within the system in the liquid phase. The high velocity near the specific boundaries of the system, the sudden disappearance of oscillations, and its dependence on evaporation are explored. This transition of classical mechanics with the outlook of quantum mechanics leaves several open questions for further investigation in the field of quantum nanofluidics.
\end{abstract}

\maketitle


\section{\label{sec:level1}Introduction}
The integrability of a physical system having classical observable to a canonically quantised system do not have a clear explanation even today \cite{hietarinta1982quantum, medvidovic2021classical}.
A reason behind this is that not all quantum parameters, such as spin has its own classical counterpart \cite{graham1984two}.
A quantum mechanical system is defined by the measurement of complete set of commuting observable and specific operators;
such system can have impeccable correlation to the classical observable.
An example of this system with classical phase-space functions could be a fluid reservoir separated by a solid interface, which creates two separate volume fractions of air and water as depicted in Figure \ref{Figure:1}a.
These fluidic systems allows mixing of the fluids through certain nanometric pores on the solid interface.
Dynamics of such systems with a single pore on the interface has already been addressed previously by us \cite{johny2021self} where the dynamics of multipore systems was unclear due to pure classical treatment.
In this paper, we present the dynamics of multipore nanofluidic systems by following a necessary transition from classical to quantum approximations.

\section{Complex self-driven nanofluidics}
\begin{figure}[htp]
\includegraphics[width=.48\textwidth]{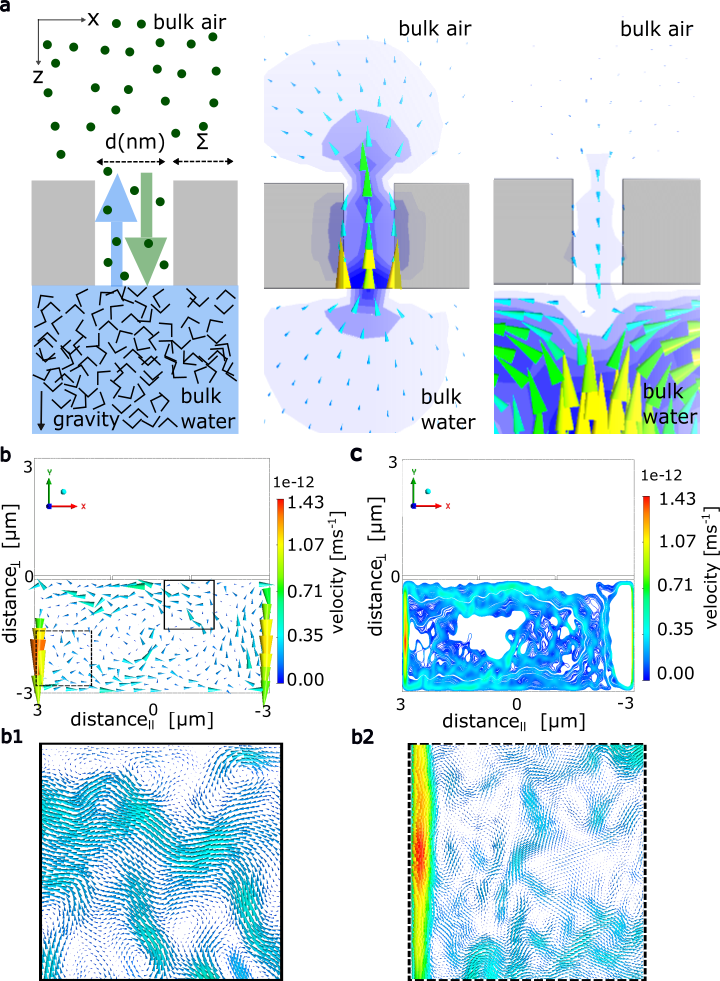}
\caption{\label{fig:1}(\textbf{a}) Overview of the phase exchange through a nanopore, and the possible along with the flow vector orientation of particles in the system.
(\textbf{b}) The velocity vectors distribution along the resultant swirls created within the water phase when no phase exchange occurs,
($d$ = 100 nm situated at $\sigma$ = 2000 nm, 4000 nm).
(\textbf{b1, b2}) Evident image of the randomness of vortices and swirls formed within a small area near the nanopore and near the walls inside liquid phase. 
(\textbf{c}) Velocity streamlines connecting 10 equidistant sample points with system.}
\label{Figure:1}
\end{figure}

The physics of nanofluidics can be empirically divided into two parts --- below and above 100 nm.  
The sub-100 nm regimes with quasi-quantum-classical approaches are quite demanding.
However, in room temperature -- water-air (liquid-gas) interface is crucial for the transpiration process in trees and, flow dynamics from such a system in classical frame of reference has significantly clear similarities to specific quantum dynamics of the same system by numerical and theoretical analysis.
Let us explore the interfacial behaviour and existence of the phases of matter inside sub-100 nm fluidic channel (where length of reservoir is much larger than the diameter of the pore) undergoing a strong differential pressure.
The flow dynamics of such systems generally falls under conventional laminar flow regime.
However, the trajectories of velocity streamlines in Figure \ref{Figure:1}b shows significant deviation from this expectation. 

The coexistence of coupled anharmonic and harmonic components in the denser volume fraction as shown in  Figure \ref{Figure:1}b will require us to use quantum mechanical approach to define its dynamics completely.
One of the critical parameters that could cause simultaneous existence of two different flow velocities in the system without any external forces can be temperature.
The peculiarities of dynamics exhibited by liquid 4-He under the lambda temperature ($\approx$2.17 K) is well known as superfluidity \cite{kapitza1938viscosity}.
 In the non-perfect BEC, the repulsive and attractive forces in the system influence superfluidity  \cite{bogoliubov1947theory}.
This is also possible in case of nonequilibrium real liquids subjected to specific operating conditions.
The coherence length of liquid 4-He diverges at the transition point and attains a steady state at  0.1 nm due to quantum coherence \cite{rojas2015superfluid}.
In such a system, the condensate moves with a very small relative velocity without any external forces which leads to the presence of super fluidity in the system.      
London \cite{tisza1938transport} considered liquid 4-He subjected to  a pressure change to exist in dual nature.
The coexistence of this dual nature with one is  being dependent on temperature in a conventional manner, and other is not being condensed even below its vapour pressure in the lowest energy state.

Hence, in the temperature range towards Lambda point the coefficient of viscosity changes from 1 $\times$ 10$^{-5}$ CGS units to 3 $\times$  10$^{-5}$ CGS units for harmonic component and nearly 0 for an-harmonic or superfluidic component \cite{tisza1938transport}. 
This is due to the anomalies in the momentum distribution function within the particles.
This impediment in a confined system at the nanometric  domain is potentially due to friction induced by the walls or particle movement.
 The ability to observe explicit quantum mechanical phenomenon in room temperature water, as in a pure quantum 4-He liquid inside nanopores is also crucial in quantum optomechanics \cite{hao2020quantum}
FEM analysis demonstrate the velocity vortices oscillating within the water phase, while the high velocity normal component is pushed towards the confined walls.
This highly anisotropic behaviour of velocity flow trajectories (connecting 10 equidistant seed points in water phase) as shown in Figure \ref{Figure:1}c along each degree of freedom, makes it complex to be validated completely using non-linear Schr\"odinger equation.
Hence, In this paper we use the Landau criterion to validate the criterion for superfluidity.

\section{\label{sec:level2}Towards quantum nano fluidics at room-temperature.}
The creation of improvised nanofluidic devices are continually looking for intrinsic explanations for observations occurring on confined systems at nanometer scale and the solutions are most likely to occur in the quantum realm.

Until the emergence of quantum thermodynamics, where we try to explain such systems using the potential of current quantum systems, some complicated phenomena at this scale were difficult to be placed merely as quantum or thermodynamic system.
One such example is a qubit system with two dimensional Hilbert space which was analysed and compared by its classical and quantum counterparts \cite{ghosh2019quantum}.
This was done by the approximation technique by Lukin et al.  \cite{lukin2000nonlinear}.

Nevertheless, a confined sub 100 nm multi-pore system exhibiting thermodynamic behaviour may seem to be solved as a classical system but its complexity and turbulent behaviour demands it to be qualitatively analysed in the quantum realm.
As researches have already been able to interpret and measure the super fluidity in nanochannels, the observations in this paper evokes the possibility of superfluidity in confined nanosystems. 
Rojas et al. \cite{rojas2015superfluid} studied the dynamics of super-fluid 4-He, coupled with super fluid Helmholtz resonance by developing a nanomechanical resonator which showed unexpected quantum turbulence.
This is due to increased dissipation within the system.
In a 9 nL system, the superfluidic mass was estimated to be  approximately 24 ng at ({T} - {T$_\lambda$}) = 2 mK and 960 ng at near 1.6 K where $T_\lambda$ is the temperature at the lambda point. 
In a confined system of 2 different phases seperated by an interfacial nanochannel, below a critical velocity the particles acts as 2 different clusters with independent physical properties.
The normal particles with higher velocity were moved towards the wall while the relatively slower superfluidic particles were dragged away from the boundaries of the confined system with respect to the reference velocity  ( reference velocity = 4.2 $\times$ $10^{-9}$ m/s).
In the nanosystem developed by Rojas et al. \cite{rojas2015superfluid}, for frequency range below 5 KHz the critical velocity was found to be 1 $\upmu$m,  which confirms the oscillation of superfluidic components in the channels and the adherence of normal components towards the walls.
Having said so, the physics behind the interaction happening in such systems is still unclear and needs to be determined.
Figure \ref{Figure:1}(b1, b2) gives a clear picture on the compexity of the transport along with velocity vectors.

\begin{figure*}[htp]
\includegraphics[width=.89\textwidth]{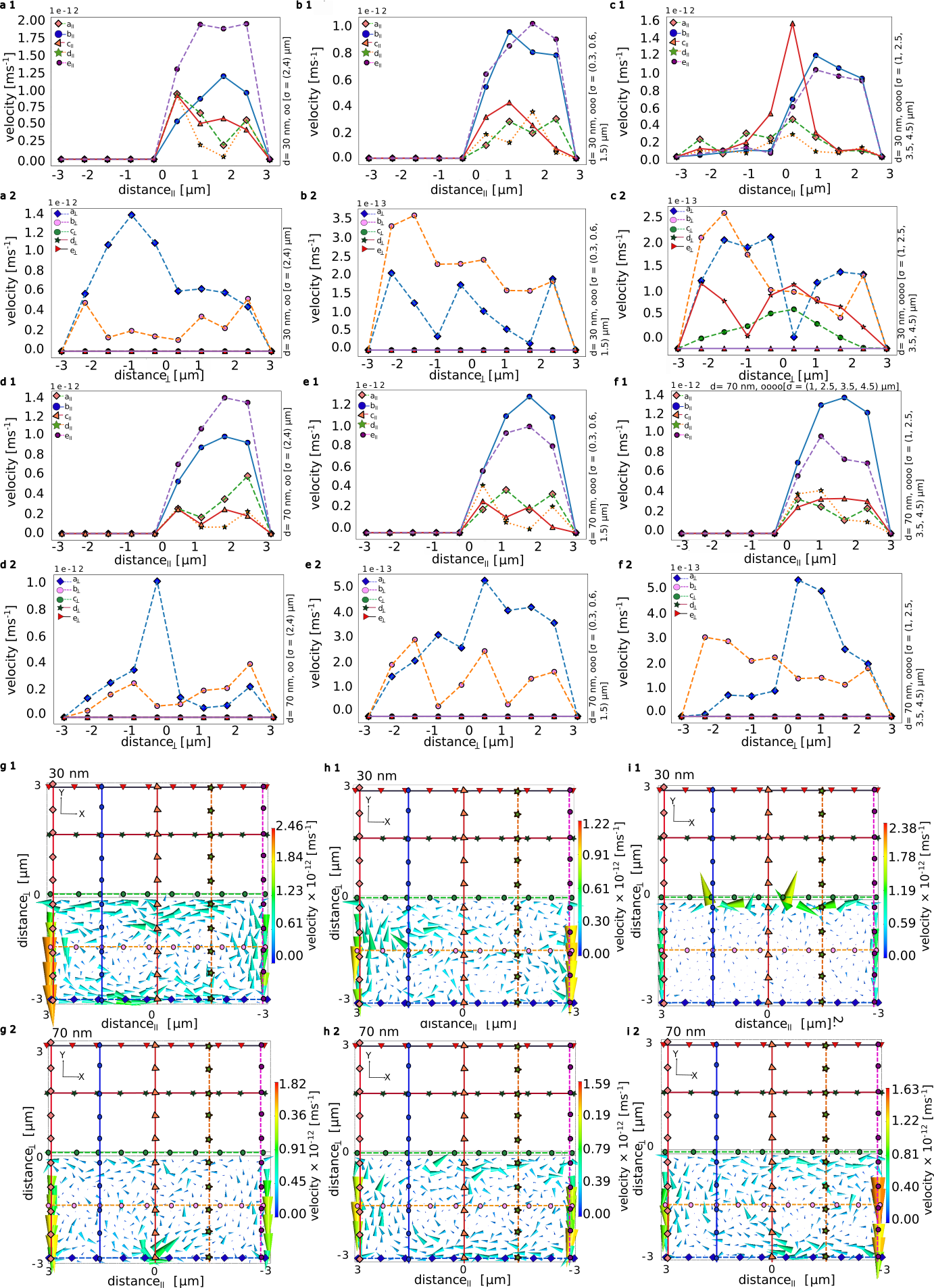}
\caption{\label{fig:2}Variation in particle velocity along the orientation of nanopore throat (distance$_{||}$), and perpendicular to the orientation of nanopore throat (distance$_{\perp}$).
(\texttt{a1, b1, c1}) Velocity variations along distance$_{||}$ for systems with two, three, four nanopores respectively for d = 30 nm.
(\texttt{a2,b2,c2})Velocity variations along distance$_{\perp}$ for systems with two, three, four nanopores respectively for d = 30 nm.
(\texttt{d1, e1, f1}) Velocity variations along distance$_{||}$ for systems with two, three, four nanopores respectively for d = 70 nm.
(\texttt{d2,e2,f2})Velocity variations along distance$_{\perp}$ for systems with two, three, four nanopores respectively for d = 30 nm.
(\texttt{g1,h1,i1}) Vector orientation of particle velocity for systems having two, three, four nanopores respectively for d = 30 nm.
(\texttt{g2,h2,i2}) Vector orientation of particle velocity for systems having two, three, four nanopores respectively for d = 70 nm.}
\end{figure*}
\section{analytical approach }
Consider the liquid phase in Figure {\ref{fig:1}b}, where the vortices are observed in the form of velocity vectors.
The vectors are highly turbulent and has varying orientation to different locations within the phase.
We are primarily interested in solving the cause of these vortices.
The reference velocity is set to {4.2 $\times$ 10$^{-9}$ m/s} and hence the observed velocity variations in Figure {\ref{fig:1}a} are very significant.
The velocity vector magnitude near to the nanopore is significantly lower (0.35 $\times$ 10$^{-12}$ m/s) than those near the walls. 
Hence, it is projected separately as in Figure {\ref{fig:1}b1} and Figure {\ref{fig:1}b2} for clear understanding.
Figure {\ref{fig:1}c} shows the velocity streamlines connecting 10 equally distant sample points within the water phase.
These seed points are  randomly distributed but are equidistant to each other to represent an overall motion of the particles in the liquid phase.
It is evident that the particle velocity near to the boundary is greater (1.43  $\times$ 10$^{-12}$ m/s) than the velocity at the center part of the liquid phase.
This may occur due to the coexistence of normal, and superfluidic components in the system.
By normal particles we mean those particles which exhibits conventional thermodynamic behavior.
Since the pressure near the nanopore at the interface is too high, the liquid-air phase transition occurs indicating the rise of temperature in the system.
Hence, if the velocity at the walls are greater than the Landau critical velocity, there is a possibility of coupling between harmonic components and anharmonic components in the {6 $\upmu$m $\times$ 3 $\upmu$m} liquid phase region (as per weak interaction  of Bose gases).
Then, this can be the visualization of the mathematical abstraction of the phenomenon of vortices in BEC.
In our previous work \cite{johny2021self}, a reverse flow was observed in certain nanocavities which suggest the direction of momentum in the opposite direction of actual flow.
This is Landau criterion for energetically unstable superfluidity.\\

At this point of our study in Figure {\ref{fig:1}a}, we assume a finite range interaction of $\zeta({\bm r_0}-{\bm r_1})$ in position space.
Then the critical velocity ${\bm v_c}$ is given by \cite{baym2012landau}
\begin{equation}
    {\bm v_c} = \Big[\Big(\frac{\bm p_c}{2m}\Big)^{2} + \frac{\rho \zeta_{\tau}}{m}\Big]^{1/2}
\end{equation}
where ${\bm p_c}$ is the critical momentum, $m$ is the mass of the particle, $\rho$ is the density, $\zeta_{\tau}$ is $\zeta(\tau \bm{p_c})$ where, $\tau$ is any integer and $\zeta(\bm{p})$ is the Fourier transform of $\zeta(\bm{r_0})$. 
When the phase velocity and group velocity of particles becomes equal, the velocity near the walls or boundary is expected to be greater than the critical velocity.
For the condensate momentum $\bm{q}$,
\begin{equation}
    {\bm{q}}^{2} > {{\bm{p_c}^2}/{4}} + {m{\rho}{\zeta_{1}}}
\end{equation}
where $\zeta_{1}$ represents the scenario where $\tau = 1$.
In system as shown in Figure {\ref{fig:1}b}, five equidistant parallel and perpendicular lines with respect to the orientation of nanopore are plotted within the water phase geometry.
Once more, 10 equidistant data points are considered along each of these lines and their velocity variations are recorded as shown in Figure {\ref{fig:2}(a-f)} to check the convergence of the solution to the criterion in equation 2.
In Figure {\ref{fig:2}a1}, the velocity magnitudes of the data points corresponding to the parallel lines to the orientation of nanopore in the order of their distribution through the geometry in two pore system ($d$ = 30 nm situated at $\sigma$ = 2000 nm, 4000 nm) is shown.
The data points on the line near to the boundary marks the highest velocity of {2 $\times$ 10$^{-12}$ m/s} in the system.
In Figure {\ref{fig:2}a2}, the velocity magnitudes of the data points corresponding to the perpendicular lines to the orientation of nanopore in the order of their distribution through the geometry in two pore system ($d$ = 30 nm situated at $\sigma$ = 2000 nm, 4000 nm) is given.
Here, the data points in the air phase drops down to zero and only two lines in the water phase shows evident deviation in flow vectors.
In Figure {\ref{fig:2}b1}, the velocity magnitudes of the data points corresponding to the parallel lines to the orientation of nanopore in the order of their distribution through the geometry in three pore system ($d$ = 30 nm situated at $\sigma$ = 2000 nm, 4000 nm) is shown.
Here, the flow vectors are prominent towards the center also, apart from boundary.
But the maximun velocity magnitude in the system has now reduced to {1 $\times$ 10$^{-12}$ m/s}.
In Figure {\ref{fig:2}b2}, the velocity magnitudes of the data points corresponding to the perpendicular lines to the orientation of nanopore in the order of their distribution through the geometry in three pore system ($d$ = 30 nm situated at $\sigma$ = 2000 nm, 4000 nm) is given.
In Figure {\ref{fig:2}c1}, the velocity magnitudes of the data points corresponding to the parallel lines to the orientation of nanopore in the order of their distribution through the geometry in four pore system ($d$ = 30 nm situated at $\sigma$ = 2000 nm, 4000 nm) is shown.
Unlike the two pore, and three pore systems of $d$ = 30 nm case, the velocity magnitude has variations in the water phase also.
Here, there is flow through the nanopore which makes significant changes in air phase also in terms of flow vectors.
The highest flow magnitude is observed through the nanopore and has magnitude of  {1.6 $\times$ 10$^{-12}$ m/s}.
In Figure {\ref{fig:2}c2}, there is significant variations in velocity components corresponding each data points.
Now, we change the pore size from 30 nm to 70 nm.
Figure {\ref{fig:2} d1} shows the flow velocity corresponding to the data points on the lines, parallel to the orientation of nanopore in $d$= 70 nm two pore system.
The maximum velocity corresponds to {1.4 $\times$ 10$^{-12}$ m/s} and is observed near the boundary.
In Figure {\ref{fig:2}d2} the maximum velocity of {1 $\times$ 10$^{-12}$ m/s} is observed at the data point located inside one of the pore.
Figure {\ref{fig:2}e1} shows the flow velocity corresponding to the data points on the lines, parallel to the orientation of nanopore in $d$= 70 nm three pore system.
In Figure {\ref{fig:2}e2}, velocity of {5 $\times$ 10$^{-13}$ m/s} is the highest of all the lines perpendicular to the orientation of nanopore in three pore system with $d$= 70 nm.
Now, the critical velocity is taken to be the velocity of the system at which highest number of data points converges regardless of change in pore size $d$.
Now, let us try to unfold our understanding on such systems in quantum realm.

\section{Outlook --- Quantum statistical approach }
Let us start from the Feynman approach towards phase transition.
At transition temperature,
consider liquid {\ch {H_2}O} in a nanochannel.
{\ch {H_2}O}, being a non linear triatomic molecule with three modes of vibrations has a frequency range between 3652 cm$^{-1}$ and 3756 cm$^{-1}$ and changes its dipole during the vibrations.
The matter waves associated with it has velocity $v$ and the wavelength $\lambda$ is defined as 
$\lambda = h/\bm{p}$.
\begin{equation}
E = \sqrt{\bm{p}^2{c^2} + {E_0}^2}
\end{equation}

\begin{equation}   
     \lambda = \frac{hc}{\sqrt{{k_e}^2 + 2{k_e}E_0}}
\end{equation}
where  $E = {k_e} + E_0$, $E_0$ is the rest mass energy, $\bm{p}$ is the momentum, and $k_e$ is the kinetic energy of the particle. 
Consider two waves $y_1$ and $y_2$ super imposing constructively such as 
$\bm{y}$ = {$\bm{y_1}$+ $\bm{y_2}$}.
then,
\begin{equation}
    \bm{y} = B[{\cos}({\omega_1}t-{k_1}x) + {\cos}({\omega_2}t-{k_2}x ]
\end{equation}
\begin{equation}
     \bm{y} = 2B [\cos({\Delta}{\omega}t-{\Delta}kx) + \cos({\omega}t-{k}x)]
\end{equation}
where $({\Delta\omega}t-{\Delta}kx)$ is the modulation in amplitude due to group velocity and $({\omega}t-{k}x)$ is the modulation of amplitude due to phase velocity, $k=$ 2{$\pi$}/{$\lambda$}, $\omega$ is the angular frequency of the wave. 
In quantum physics, the velocity of particle is referred to its group velocity and hence in a system of particles with different group velocities, the rate of difference in phases can determine the rate of superfluidity in the system with respect to its reference velocity.
Their velocities will be out of phase at the zone boundaries and in the center by which the initial coherent state of the particles are disturbed.
It is expected that when the minimum velocity in our system becomes equal to the Landau critical velocity, the phase velocity becomes equal to the group velocity\cite{baym2012landau}.

\begin{equation}
    \bm v_g = \bm v_p + k(d{\bm v_p} / dk) 
\end{equation}
 which leads to the dispersion relation,
\begin{equation}
    \bm v_g = \bm v_p - \lambda(d{\bm v_p} / d\lambda) 
\end{equation}

This relates to the energy exchange within the strong and weak coupling which is applicable at the quantum-classical boundaries of the system \cite{reithmaier2004strong}.
Lukin, in his work on ultra slow single photons showed that one among two weak waves of energy $h\nu$ can modify the properties of one component wave resulting in a phase shift \cite{mebrahtu2012quantum}.
The quantum entanglement thus formed causes super positions and quantum interference.
A mathematical model to understand such non linear quantum systems was done by \cite{erdHos2007derivation}. 
In his work, he considered the nonlinear Schr\"odinger Gross-Pitaevskii equation to understand the quantum dynamics of a cubical system, where the maximum particle interaction scale is the length of the system.
At this point, it is vital to determine if the system is in equilibrium or whether the observation has a probability of decaying over time.
The nonlinear dispersive effects on the wave vector cancel out the steepening effect of waves in linear approximation, and the system achieves an equilibrium state, according to Tsuzuki\cite{tsuzuki1971nonlinear}.
In such a system, where the ground state scattering length has a strong dependence on the repulsive interactions, Gross-Pitaevskii equation could pave light to the dynamics of the particles within the system.

This may be complex problem due to the highly anisotropic nature of the particles along each axis, and would create a quantum confinement of many body system.

\begin{figure*}[htp]
\includegraphics[width=0.8\textwidth]{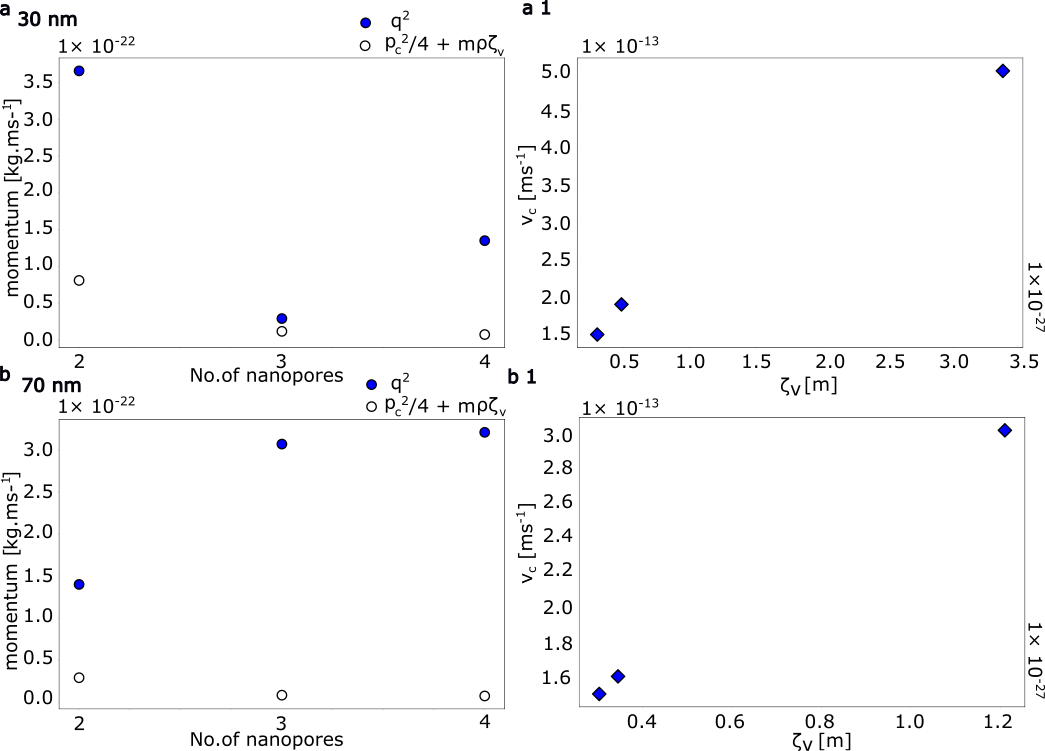}
\caption{\label{fig:3}Analytical validation of the landau criterion in 30 nm and 70 nm nanpore systems and the dependency of finite range interaction with critical velocity.
(\textbf{a}) Validation of the Landau criterion for d = 30 nm .
(\textbf{b}) Validation of Landau criterion for d = 70 nm. 
(\textbf{a1}) Change in critical velocity of a system with the increase in finite  range interaction in posistion space for d = 30 nm. 
(\textbf{b1}) Change in critical velocity of a system with the increase in inite  range interaction in position space for d = 70 nm.} 
\end{figure*}

Let us start with the identity of the particles in the 2D continuum of states in the system.
For the particles to obey BEC, they should obey the Bose Einstien statistics, What if these particles are Fermions as non-Hermitian superfluidity is also a possibility here?
The answer to this question lies in the observed vortex pairs of the system which shows that these particles are of type-1 bosons as in \cite{kohn1970two}.
As 4-He atom which is considered to be of such kind, bound pair of spin half particles as seen in \cite{johny2021self} can be considered to have characteristics as that of bosons.
Hence we begin our journey with the  method of second quantization in the context of Quantum field theory.
As it defines many body operators which are crucial in superconductivity and superfluidity \cite{kanazawa2021non} in terms of creation and annihilation operators.
 We are aware of the switching mechanism happening in the system which is unclear in the classical domain.
 Let us try to explain the system in this context, as every measurements are associated with creation and annihilation operators \cite{melkikh2015nonlinearity}. 
 If the system has $N$ number of single particle state
 $| \phi_b\rangle$ (orthonormal and complete, $b$ is the index associated with all the quantum numbers)
, the action of creation operator creates a particle in the single particle state and if already a particle is present, then the effective action of the operator becomes zero.
 As for an annihilation operator, if a particle is present in a single particle state, it annihilates it.
 If the state is already empty, there is no effective action.
 Then the scattering of a particle from state $| \phi_b\rangle$ to $| \phi_ {\chi}\rangle$ is due to the annihilation operator followed by creation in state $| \phi_ {\chi}\rangle$. 
 This justifies to the dispersion relation from equation 8.
 The Hamiltonian of such system can be given as
 \begin{equation}
 H_0 = \sum_{{b}{\chi}} \langle{\phi_{\chi}} {|h|} {\phi_{b}}\rangle {{\varphi_{\chi}}^{\dagger} {\varphi_{b}}}
 \end{equation}
 such that h(i) is an operator dependent on the space cordinates of the particle \cite{jishi2013feynman}.
 where $\varphi^{\dagger}$ is the creation operator and $\varphi$ is the annihilation operator.
 Consider the basis state associated with spin $\Upsilon$ and wave number $k$,  the kinetic energy of the system can be expressed as,
 \begin{equation}
 {K_e} = \sum^{N}_{i=1}\big({\frac{-\hbar^2}{2m}\big)} {\nabla^2_{i}}
 \end{equation}
 \begin{equation}
 \phi{_{k\Upsilon}}(r) = \big\langle {r|k\Upsilon}\big\rangle = \frac{1}{\sqrt{V}} \exp({i k r})|\Upsilon\big\rangle
 \end{equation}
 where in 2D confinement, $k_x$, $k_y$ $=$ 0, ${\pm}\frac{2\pi}{L}$, ${\pm}\frac{4\pi}{L}$...\\
 \begin{equation*}
                 {K_e} = \sum_{k\Upsilon}{\sum_{k^{1}\Upsilon^{1}}} {\big\langle{{k^1}{\Upsilon^1}|\frac{-\hbar^2}{2m}{\nabla^2}|{k\Upsilon}\big\rangle}}{\varphi{^\dagger}{_{{k^1}{\Upsilon^1}}}}{ \varphi_{k\Upsilon}}\\
     \end{equation*}
 \begin{equation*}
 {K_e} = ({\hbar^2{k^2}}/2m) \varphi^\dagger{_{k\Upsilon}} \varphi_{k\Upsilon}
 \end{equation*}

 As confirmed from the analytical approach, as the delta function can express the particle number density at a position which is closely associated with the creation-- annihilation operators.
 But this is not adequate to give the number of particles in the system.
 Even though there are no significant physical changes in both air-water phases, particle exchange and energy exchange are evident enough to consider it as a member of the grand canonical ensemble.
 Then the ensemble average of total number of particles is obtained
as the Bose-Einstien quantum distribution function (\begin{math}
{n_i} = \frac{1}{\exp{\beta(\epsilon_i - Q)} - 1}
\end{math}).
where $Q$ is the chemical potential of the system, $\beta = \frac{1}{kT}$, $\epsilon_i$ is the energy state corresponding to the coordinate $i$.

\section{DISCUSSION}
The observed highy anisotropic and non linear behaviour of particles for a laminar Reynolds number would be a tedious problem to encounter.
For convenience, we plotted the velocity variation along parallel and perpendicular lines along the orientation of the nanopores as illustrated in figure.2.
Systems having $d =$ 30 nm, and $d =$ 70 nm were simulated using ANSYS FLUENT 2021 R1 version,  where $d$ is the nanopore width.
For each $d$, pore systems with two, three, and four nanopores were simulated separately to resolve their dynamics more efficiently and precisely.
Fig \ref{fig:2}a1, figure {\ref{fig:2}b 1}, figure {\ref{fig:2}c1} shows the velocity variations over the seed points in systems having 2, 3, 4 nanopores respectively for $d =$ 30 nm.

Similarly, Figure {2d1}, Figure {2e1}, Figure {2f1} shows the velocity variations over the seed points in systems having 2, 3, 4 nanopores respectively for d = 70 nm.
Even though the orientation and direction of the velocity vectors of these systems are random and non-linear (Figure {\ref{fig:2}(g - i)}, most of the individual data points has a velocity near to 
$0.2\times10^{-12}$~ms$^{-1}$
regardless of their position which was considered to be the critical velocity.
The system with most transport through the nanopore was the system with $d =$ 30nm and hence there were no vortices within the water phase and shows relatively high deviation in velocity profile (Figure 2c1). 
This indicates the presence of transport through the nanopore and the absence of oscillating condensate within the water phase due to phase transition and diffusion.
In all other systems, the velocity of \texttt{line $c_{||}$} which indicates the flow through the nanopore 
was significantly lower, implying that there was very less
flow to the air domain, which supports the objective of this study.
Extreme flow velocity may clearly be seen at the container's extreme opposing ends, which are marked by \texttt{lines b$_{||}$, e$_{||}$}.
The velocity tends to be similar as you get closer to the centre of the liquid domain, where the vertices develop.
According to the Landau criterion of superfluidity, the phase velocity becomes equal to the group velocity at the critical velocity and, the ninth sample point of most of the a harmonic components has a velocity of $2.4 \times 10^{-12}$ m/s regardless of its position.
Now, it is evident that the total number of particles within the water phase can be divided into two groups based on their velocity, location and the amount of turbulent characteristics they exhibit.
The relatively low velocity components are an harmonic in nature and forms the condensate, which oscillates withing the confinement of the container.
These oscillations are again dependent on the forces of interaction within the interacting range and also coupled with the system's normal harmonic components that are pushed against the walls.
In general for $d$ = 30 nm and $d$ = 70 nm, the square of the condensate momentum were higher than 
\begin{math}
\frac{{\bm{p_c}}^2}{4} + m\rho\zeta_{\tau}
\end{math} for all two, three, four, pore systems respectively as shown in figure \ref{fig:3}a and figure \ref{fig:3}b.
These results lead us to the objective of this paper with the possibility of further investigation in such systems theoretically and through real time experimentation.
The graphs in figure \ref{fig:3}c and figure \ref{fig:3}d reveals a lot about critical velocity's behaviour as the physical state of the system changes.
It also has a linear relationship with the number of pores in the system and, it is highly dependent on the surface to area ratio within the interaction range.\par

By solving the equations of motion of a classical system, we demonstrate instantaneous dynamic variables, which can predict the non-trivial flow behaviour of a nanofluidic liquid-air interfacial system. 
However, in quantum mechanics, the equation of motion becomes the variation of expectation values with time for the state vector in the abstract Hilbert space. 
The defects in our systems are the nanofluidic pore in the solid interface between the air-liquid phases. 
The two major points to be noted are, the change in dynamics with vanishing potential strongly depends on the defect being created in the system which satisfied the Landau criterion.
They are dependent on the finite-short range interactions occurring within the liquid phase.
Our results evident that the critical velocities of 30 nm and 70 nm increase with the finite short-range interaction strength due to the defects, which is in agreement with Ianeselli et al. \cite{ianeselli2006beyond}.
We emphasise that numerical insights to the dissipative term in these systems are crucial for solving nonlinear Schr\"odinger equation and is essential for the transition of such system to a quantum mechanical system.

\textit{Acknowledgments}: The VJ and SG thank Honeywell CSR fund awarded to the International Center for Nano Devices.
SG thanks the German Research Foundation and Isaac Newton Trust for funding his research. 
SG is grateful to Professor Natalia Berloff for many productive discussions.

\appendix
\section{Notations}
\begin{table}[h!]
\begin{tabular}{ l  l }
\hline
$B$ & Wave amplitude\\
$d$ & Diameter of the Nanopore\\
$E$ & Energy\\
$\eta$ & Viscosity of liquid\\
$\epsilon_i$ & QM energy state corresponding to the coordinate $i$\\
$\gamma$ & Phase volume fraction in the multiphase flow\\
$\Gamma$ & Cross sectional area\\
$h(i)$ & Quantum mechanical operator\\
$H_o$ & Hamiltonian of the system\\$\kappa$ & Phase variable\\
$k$ & Wave number\\
$k_B$ & Boltzmann constant\\
$L$ & Flow length of instability\\
$M$ & Mass transport through the nanopore\\
$N$ &  Total number of particles in the system\\
$P$ & Pressure\\
$\bm{p_c}$ & Critical momentum\\
$\bm{q}$ & Condensate momentum\\
$r$ & Radius of nanofluidic pore\\
$\rho$ & Phase density\\
$\bm{r}$ & Position vector\\
$\sigma$ & Position of nanopore from the reference edge \\
$\tau$ & Constant corresponding to equation 1 \\$\bm v_p$ & Phase velocity\\
 $\bm v_g$ & Group velocity\\
$\bm v_c$ & Critical velocity\\
$W$ & Energy corresponding to work function of evaporation\\
$\zeta$ & Finite range interactions between particles\\
$\phi_b$ & Quantum state of system before interaction\\$
\phi_{\chi}$ & Quantum state of system after interaction\\
$\varphi^\dagger$ & Creation operator\\
$\varphi$ & Annihilation operator\\
$\Upsilon$ & Spin\\
$K_e$ & Kinetic energy of the quantum system\\
$Q$ & Chemical potential\\
\hline
\end{tabular}
\end{table}

\bibliography{ref}

\end{document}